\documentstyle[aps,prl,array,multicol,psfig]{revtex}

\def\wide#1#2{
\end{multicols}
\widetext
\noindent
\if#1t
\else
    \raisebox{9pt}[0in][0.0in]
    {$\rule{3.4in}{0.4pt}\rule{0.4pt}{6pt}$\hspace{3.6in}}
\fi
#2
\if#1b
\else
    \raisebox{-9pt}[0in][0.0in]
    {\hspace{3.55in}$\rule{0.4pt}{6pt}\rule[6pt]{3.5in}{0.4pt}$}
\fi
\begin{multicols}{2}
\noindent
}

\begin{document}

\title{Hydrodynamics of the quantum Hall smectics}

\author{Michael M. Fogler}

\address{School of Natural Sciences, Institute for Advanced Study,
Olden Lane, Princeton, New Jersey 08540}

\author{Valerii M. Vinokur}

\address{Material Science Division, Argonne National Laboratory,
Argonne, Illinois 60439}

\date{
\today\
}

\maketitle


\vspace{-0.3in}

\begin{abstract}

We propose a dynamical theory of the stripe phase arising in a
two-dimensional electron liquid near half-integral fillings of high
Landau levels. The system is modelled as a novel type of a smectic
liquid crystal with the Lorentz force dominated dynamics. We calculate
the structure factor, the dispersion relation of the collective modes,
and their intrinsic attenuation rate. We show that thermal
fluctuations cause a strong power-law renormalization of the elastic
and dissipative parameters familiar from the conventional smectics but
with different dynamical scaling exponents.

\end{abstract}

\begin{multicols}{2}

Planar arrays of interacting lines or stripes have become a paradigm
for many different physical systems, including domain walls in
magnets, layered superconductors, biophysical systems, liquid
crystals, and charge density waves (CDW)~\cite{Seul}. A unidirectional
CDW or the stripe phase was also predicted~\cite{Koulakov_96} to form
in GaAs two-dimensional (2D) electron systems when the occupation of
the third or a higher Landau level is close to
$\frac12$~\cite{Fogler_96}. This prediction was attested by a recent
discovery of a dramatic magnetotransport
anisotropy~\cite{Experiment_QHE} in the indicated range of filling
factors. The easy (low resistance) and hard (high resistance) current
directions are thought to be along and across the stripes,
respectively. In order to have a better foundation for the
magnetotransport studies, one has to first understand the intrinsic
dynamics of the stripe phase, without intervening disorder
effects~\cite{Transport_theory}. Recent
investigations~\cite{Fradkin_99,Fertig_99,MacDonald_00} revolved
around the instability of the stripes against the $2 k_F$-modulation
along the direction of the translational order, which would transform
the system into a highly anisotropic Wigner crystal. According to
MacDonald and Fisher~\cite{MacDonald_00}, such an instability appears
only at temperatures $T$ below $1\, {\rm mK}$. Here we would like to
address another interesting regime of relatively high $T$ where the $2
k_F$ modulation is absent and moreover, the sole periodic modulation
is due to the main CDW harmonic with wave vector $q_0$. In classical
terms, the electron density at the topmost Landau level is of the form
$n({\bf r}, t) + {\rm Re}\,\Psi({\bf r}, t)$, where $\Psi = |\Psi|
e^{i q_0 (x - u)}$ is the CDW order parameter. (We chose the $\hat{\bf
x}$-direction to be perpendicular to the stripes). Although
oversimplified, this classical picture correctly identifies the
low-frequency long-wavelength degrees of freedom: the coarse-grained
component $n$ of the total density, and the CDW phase (or the
displacement field) $u$. It is natural to expect that at sufficiently
small $\omega$ and $q$ the dynamics of $u$ and $n$ is governed by a
certain hydrodynamic theory. The formulation of such a theory is the
subject of this Letter.

We begin by writing down an effective Hamiltonian, from which we
derive the hydrodynamic equations of motion, the spectrum of the
collective modes, and the structure factor. Next, we study the
renormalization of the bare parameters of the theory by thermal
fluctuations. Our results can be verified by measuring the
transmission of the micro- and surface acoustic waves as well as
inelastic light scattering techniques provided the samples are of high
enough mobility so that the disorder effects, which we ignore here,
are not important.

{\it Effective Hamiltonian\/}.---
Our first step is to construct an effective Hamiltonian for $n$ and
$u$, which is (a) rotationally and translationally invariant
and (b) reflects the specific properties of the system, such as the
long-range Coulomb interaction. Very useful in this process is the
similarity~\cite{Fradkin_99} to smectic liquid
crystals~\cite{deGennes}. The resultant form of the effective
Hamiltonian is as follows:
\begin{eqnarray}
H = \frac12 \int d^2 r [Y E^2(u) + E(u) \Delta
  + K (\bbox{\nabla}^2 u)^2 &&
\nonumber\\
\mbox{} + \chi^{-1} \delta n^2 + \Phi n + 2 C E(u) \delta\rho + \rho v^2],&&
\label{H}
\end{eqnarray}
where $E(u) = \nabla_x u - \frac12 (\bbox{\nabla} u)^2$ in the usual
rotationally invariant strain~\cite{deGennes}, $Y$ and $K$
are the compression and bending elastic moduli, $\chi$ is the
compressibility~\cite{Comment_on_exchange}, $\Delta$ is an
auxillary parameter (counter-term), needed to guarantee the condition
$\langle \nabla_x u \rangle = 0$, $\rho = m n$ is the mass density,
$\delta \rho = \rho - \rho_0$ is its deviation from the equilibrium
value $\rho_0 = m n_0$. $\Phi({\bf r}) = \int d^2 r' \delta n({\bf
r}') U({\bf r} - {\bf r}')$, is the electrostatic (Hartree) potential,
with $U(r) = e^2 / \kappa r$ at large $r$ (see more details
in~\cite{Koulakov_96}). The penultimate term in Eq.~(\ref{H})
accounts for the dependence of the CDW periodicity on $n_0$, with $C =
(Y / m) \partial \ln q_0 / \partial n_0$. Finally, the last term is
the correction to the kinetic energy, with $v$ being the velocity of
the electron fluid. This term is mainly a bookkeeping device: it
vanishes after the projection on a single (topmost) Landau level but
enables us to derive the equations of motion via the standard
Poisson-bracket method~\cite{Mazenko_83,Kats_Lebedev}.

{\it Equations of motion\/}.--- The hydrodynamic fluctuations of our system
are governed by the equations
\begin{eqnarray}
F_u &\equiv& \rho_0 \frac{\dot{W} + ({\bf v} \bbox{\nabla}) W}{|\nabla W|}
 - \lambda_p |\nabla W| \frac{\delta H}{\delta u}
 = \zeta_u,
\label{F_u}\\
F_j &\equiv& \frac{\partial (\rho v_j)}{\partial t} + \nabla_j
(\rho v_k v_j) - \rho \omega_c \varepsilon_{j k} v_k
 + \nabla_j W \frac{\delta H}{\delta u}
\nonumber\\
\mbox{} &+& L_{j k}(i \bbox{\nabla}) v_k + \rho \nabla_k \left(
\chi^{-1} \delta\rho + \Phi + C E\right) = \zeta_j,
\label{F_v}
\end{eqnarray}
and the continuity equation $\dot{\rho} + \bbox{\nabla}(\rho {\bf v}) =
0$. The notations here are as follows: $W \equiv x - u$, $\varepsilon_{j
k}$ is the unit antisymmetric tensor, and $\omega_c = e B / m c$ is the
cyclotron frequency. Functions $L_{j k}({\bf q})$ describe dissipation.
As in 3D case~\cite{deGennes}, they can be parametrized by viscosities
$\eta_i$,
\begin{eqnarray}
L_{j k}({\bf q}) &=& \eta_4 q_j q_k + \delta_{j k} \eta_3 q_x^2
\nonumber\\
\mbox{} &+& \delta_{j x} \delta_{k x} [\eta_3 q^2 + (\eta_1 - 4 \eta_3
+ \eta_4 - 2 \eta_5) q_x^2]
\nonumber\\
\mbox{} &+& (\eta_3 - \eta_4 + \eta_5) q_x
(q_j \delta_{k x} + q_k \delta_{j x}).
\label{L}
\end{eqnarray}
(There is no analog of $\eta_2$ in 2D). One more dissipative
coefficient, $\lambda_p$, [Eq.~(\ref{F_u})] describes the permeation,
i.e., the mass transport across the stripes. At relatively high $T$,
such that $|\Psi| \lesssim n_0$, $\lambda_p \sim (\eta_3 q_0^2 / \rho_0
\omega_c^2) (n_0 / |\Psi|)^2$. Finally, $\zeta_a$'s in Eqs.~(\ref{F_u})
and (\ref{F_v}) are the Langevin noises, which satisfy the
fluctuation-dissipation theorem,
\begin{eqnarray}
&& \langle \zeta_u(1) \zeta_k(2) \rangle = 0,
   \quad 1(2) \equiv \{{\bf r}_{1(2)}, t_{1(2)}\},
\\
&& \langle \zeta_u(1) \zeta_u(2) \rangle = 2 k_B T \lambda_p \rho_0
\delta(1 - 2),
\\
&& \langle \zeta_j(1) \zeta_k(2) \rangle = 2 k_B T
L_{j k}(i \bbox{\nabla}) \delta(1 - 2).
\end{eqnarray}
%

The main difference of our model from the conventional smectics is the
presence of the strong magnetic field. As we will see below, it
drastically changes both the linearized and nonlinear dynamics of the
system.

{\it Field-theory description\/}.--- Our ultimate goal is to
calculate various correlation functions, e.g., the dynamical
structure factor,
\begin{equation}
%
S({\bf q}, \omega) = (N_0 m^2)^{-1} \int d t e^{i \omega t}
\langle \rho_{\bf q}(t) \rho_{-\bf q}(0)\rangle,
\label{S_def}
\end{equation}
where $N_0$ is the total number of electrons at the topmost Landau
level, and $\langle \ldots\rangle$ stands for the thermal averaging, or
equivalently, the averaging over the Langevin noise. The latter can be
done in a systematic way by means of the Martin-Siggia-Rose (MSR)
formalism~\cite{MSR}, whose field-theoretic version was
previously applied to the conventional smectics by Kats and
Lebedev~\cite{Kats_Lebedev}.

The implementation of the MSR method begins with enforcing the equations
of motion by $\delta$-function-type weight factors, $\delta(F_a -
\zeta_a)$, in the path integral over $F_a$. These $\delta$-functions are
then represented by integrals of $\exp[i p_a (F_a - \zeta_a)]$ over
auxillary dynamical variables $p_a$, which enables one
to average over the Gaussian noise $\zeta_a$. Finally, one
changes the integration variables from $F_a$ to $V_i \equiv \rho v_i /
\rho_0$ and $u$. If the denote by $F^{nl}_a$ the nonliner terms in
$F_a$, and introduce vectors $\bbox{\phi}^\dagger = \{V^\dagger,
P^\dagger\}$, $V^\dagger = \{u, V_x, V_y\}$, $P^\dagger = \{p_u, p_x,
p_y\}$, then the resultant action becomes
\begin{eqnarray}
&& A = -i \int d t \int d^2 r p_a F^{nl}_a(u, v) - \ln J + A_0,
\label{A}\\
&& A_0 = \frac12 \int \frac{d^2 q}{(2 \pi)^2} \int \frac{d \omega}{2 \pi}
\bbox{\phi}_{{\bf q} \omega}^\dagger {\bf G}_0^{-1}({\bf q}, \omega)
\bbox{\phi}_{{\bf q} \omega},
\label{A_0}
\end{eqnarray}
where $J = |\det \partial F_a / \partial V_b|$ is the Jacobian and ${\bf G}_0$
is the (bare) propagator of the following block-matrix form
\begin{equation}
 {\bf G}_0 = \left[\begin{array}{cc} 
{\bf G}_{V V} & -i {\bf G}_{V P}\\
-i {\bf G}_{V P}^\dagger & {\bf 0}
\end{array}\right].
\label{G_0}
\end{equation}
Two particular components of ${\bf G}_0$ , $G_{u u}$ and $G_{u p_x}$,
\begin{eqnarray}
&& G_{u p_x}({\bf q}, \omega) = \displaystyle
\frac{-i s^2}{Q({\bf q}) s^2 - i \omega \tilde\nu({\bf q}) -
 \omega^2 \frac{\omega_c^2}{\omega_p^2(q)}},
\label{G_upx}\\
&& G_{u u}({\bf q}, \omega) =  \displaystyle \frac{k_B T}{\rho_0 \omega}
[G_{u p_x}({\bf q}, \omega) - G_{u p_x}({\bf q}, -\omega)],
\label{G_uu}
\end{eqnarray}
will play an important role in the later discussion. Here we introduced the
notations $Q({\bf q}) = (Y q_x^2 + K q^4) / \rho_0$, $\omega_p^2(q) =
n_0 q^2 [U(q) + \chi^{-1}] / m$, $\tilde\nu_i = \eta_i q^2 / \rho_0$, $c
= q_x / q = \cos\theta$, $s = q_y / q = \sin\theta$, and
\begin{equation}
\tilde\nu({\bf q}) = \tilde\nu_3 + c^2 s^2
(\tilde\nu_1 - 4 \tilde\nu_3 + \tilde\nu_4 - 2 \tilde\nu_5).
\label{nu}
\end{equation}
Equations (\ref{G_upx}) and (\ref{G_uu}) are obtained by neglecting
$\omega$, $\lambda_p$, $\tilde\nu_i$, and $C$ compared to quantities
proportional to ``large'' frequencies $\omega_c$ and $\omega_p$.

{\it Harmonic theory\/}.--- If only the quadratic part $A_0$ of the full
action is retained, then the correlators of the velocity fields
$V_a$ are given by the components $G_{V_a V_b}$ of ${\bf G}_0$
and can be found after straightforward albeit tedious
algebra. The continuity equation enables us to relate these correlators
to the structure factor. This way we get
\begin{equation}
S({\bf q}, \omega) = \frac{2}{m \omega} {\rm Im}\,
\frac{k_B T q^2 (Q s^2 - i \omega \tilde\nu)}
{Q \omega_p^2 (s^2 + \lambda_p \tilde\nu) - i \omega \omega_c^2 \alpha
 - \omega^2 \omega_c^2},
\label{S}
\end{equation}
for $\omega \ll \omega_c$ and, in particular,
\begin{equation}
S({\bf q}, 0) = 2 k_B T \frac{q^2}{m \omega_p^4}
\frac{\omega_c^2 \lambda_p + c^2 \tilde\nu_3 + s^2 \tilde\nu_4}
{s^2 + \lambda_p \tilde\nu}.
\label{S_0}
\end{equation}

The poles of $S({\bf q}, \omega)$ correspond to the collective modes.
Two of them (magnetophonons) are gapless:
\begin{equation}
\omega_m({\bf q}) \simeq \sin 2 \theta
 \left(\frac{Y}{\rho_0}\right)^{1/2} \frac{\omega_p(q) q}{2 \omega_c}
 - \frac{i}{2} \alpha({\bf q})
\label{omega_m}
\end{equation}
and its counterpart $\omega = -\omega_m^*({\bf q})$. Their attenuation
rate $\alpha({\bf q})$ is given by
\begin{equation}
\displaystyle \alpha({\bf q}) = \lambda_p Q + 
\omega_c^{-2}
[\omega_p^2(q)\tilde\nu + Q(\tilde\nu_3 c^2 + \tilde\nu_4 s^2)],
\label{alpha}
\end{equation}
up to terms proportional to $C$ and higher powers of $\tilde\nu_i$ and
$\lambda_p$. The other two collective modes (magnetoplasmons) have a large gap
$\omega_c$. Thus, the number of the hydrodynamic (gapless) modes
coincides with the number of the hydrodynamic variables ($n$ and $u$) as
it should~\cite{Comment_on_v}.

For Coulomb interaction $\omega_p(q) \propto q^{1/2}$;
hence, within the harmonic theory the magnetophonons have
$\omega_m({\bf q}) \propto q^{3/2}$ dispersion, similar to that of the
Wigner crystal~\cite{Bonsall_77} but with the $\theta$-dependent
prefactor. Propagating magnetophonon modes exist as long as $\theta$ is
not too small so that ${\rm Re}\, \omega_m({\bf q}) \gg \alpha({\bf q})$
in Eq.~(\ref{omega_m}); otherwise, they are replaced by the
two overdamped modes:
%
\[
\omega_{\rm fast}({\bf q}) \simeq -i \alpha({\bf q}),\quad
\omega_{\rm slow}({\bf q}) \simeq -\frac{i}{\alpha({\bf q})}
\left(\frac{q_x q_y}{q} \frac{\omega_p}{\omega_c}\right)^2 \frac{Y}{\rho_0}.
\]
%

Below we will see that at small enough $q$ the magnetophonon dispersion
relation and their damping are significantly modified by the
anharmonisms.

\begin{figure}
\centerline{
\psfig{file=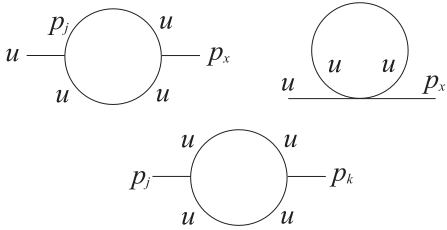,width=2.3in,bbllx=171pt,bblly=539pt,bburx=421pt,%
bbury=670pt}
}
\vspace{0.1in}
\setlength{\columnwidth}{3.2in}
\centerline{\caption{Self-energy diagrams.
\label{Sigma_fig}
}}
\end{figure}

{\it Renormalization\/}. Our next step is to calculate the propagator
${\bf G} = ({\bf G}_0^{-1} + \bbox{\Sigma})^{-1}$ of the full theory,
Eq.~(\ref{A}), treating the previously ignored nonlinear terms as
perturbations. We expect that ${\bf G}$ has the same form as ${\bf G}_0$
but with $Y$, $K$, and other parameters replaced by $\omega$ and
$q$-dependent (renormalized) values $Y^R({\bf q}, \omega)$, $K^R({\bf
q}, \omega)$, {\it etc.\/} The perturbative corrections to $Y$, $K$, and
$C$ are determined by the self-energy component $\Sigma_{u p_x}({\bf q},
\omega)$. Similarly, the corrections to $\nu_i$'s are determined by
$\Sigma_{p_j p_k}({\bf q}, \omega)$. To the lowest order in $T$ they are
given by the diagrams shown in Fig.~\ref{Sigma_fig}.

%
%

Doing the power counting with the help of Eqs.~(\ref{G_upx}),
(\ref{G_uu}), and the relation $k_x \sim k^2 (K / Y)^{1/2}$, where $k$
is the loop momentum, we quickly discover that $\Sigma_{u p_x}$ and
$\Sigma_{p_j p_k}$ are infrared-divergent. To resolve this problem, we
utilize the renormalization group (RG) procedure formulated in $d = 3 -
\epsilon$ spatial dimensions [one $x$-dimension plus $(d - 1)$
``$\perp$''-dimensions]. Each step of our RG transformation consists of
integrating out the cylindrical shell $\Lambda e^{-l} < k_\perp <
\Lambda$ in the loop diagrams of Fig.~\ref{Sigma_fig}, followed by
rescaling of the momenta $k = k^\prime e^{-l}$ to restore the
ultraviolet cutoff $\Lambda \sim q_0$ on $k_\perp$. It is convenient to
rescale the $u$-field as well $u({\bf r}) = u^\prime({\bf r}^\prime)
e^l$ to preserve the structure of $E(u)$. We did not find it necessary
to rescale the frequencies or impose a cutoff on
$k_x$~\cite{Radzihovsky_99}. The one-loop RG equations are given by
\begin{eqnarray}
&& \frac{d}{d l} Y = d Y - \frac{g}{8} Y,\quad
   \frac{d}{d l} C = d C - \frac{g}{8} C,
\label{Y_flow}\\
&& \frac{d}{d l} K = (d - 2) K + \frac{g}{8 (d - 1)} K,
\label{K_flow}\\
&& \frac{d}{d l} g = (3 - d) g - \frac{d + 2}{16 (d - 1)} g^2,
\label{g_flow}\\
&& \frac{d}{d l} \nu_3^2 = 2 d \nu_3^2 + \frac{g}{2 (d - 1)}
\nonumber\\
&&\quad\quad\quad\mbox{} \times \left[
\frac{\omega_c^2}{\omega_p^2(\Lambda e^{-l})}
\frac{K(l)}{\rho_0} e^{(d + 2) l} + \frac{\nu_3^2}{4}\right],
\label{nu_3_flow}
\end{eqnarray}
where $g \equiv k_B T Y^{1 / 2} K^{-3 / 2} \Lambda^{d - 3} / S_{d - 1}$
is the dimensionless coupling constant, with $S_{d - 1} = (4
\pi)^{\frac{d - 1}{2}} \Gamma\left(\frac{d - 1}{2}\right) / 2$.
Equations for other viscosities are similar to~(\ref{nu_3_flow}) and
are not shown.

The solutions of the RG equations~(\ref{Y_flow}--\ref{nu_3_flow})
are as follows:
\wide{m}{
\begin{eqnarray}
&&\displaystyle g(l) = g_* + \frac{g_0 - g_*}{D},\quad
D \equiv 1 + \frac{g_0}{g_*} [e^{(3 - d) l} - 1],
\quad
 \nu_3^2(l) = e^{2 d l} \left[\nu_{3}^2(0)
+ \frac{1}{2 (d - 1)} \frac{\omega_c^2}{\omega_p^2(\Lambda e^{-l})}
\frac{K_0}{\rho_0}\right] D^{\frac{2}{(d + 2)}},
\label{g_nu_3_l}\\
&& Y(l) = Y_0 e^{d l} D^{-\gamma_Y},\quad C(l) = C_0 e^{d l} D^{-\gamma_Y},
\quad \gamma_Y = 2 \frac{d - 1}{d + 2},\quad
K(l) = K_0 e^{(d - 2) l} D^{\gamma_K},\quad
\gamma_K = \frac{2}{d + 2}.
\label{Y_K_l}
\end{eqnarray}
}

As one can see, for small $\epsilon = 3 - d$, $g$ flows to a weak
coupling fixed point $g_* = 16 (3 - d) (d - 1) / (d + 2)$, justifying
our one-loop RG. The renormalized values of the parameters of the
harmonic theory are found by rescaling back to the original coordinates,
$Y^R(\omega, {\bf q}) = Y(l) e^{-d l}$, $K^R(\omega, {\bf q}) = K(l)
e^{-(d - 2) l}$, $\nu_3^R(\omega, {\bf q}) = \nu_3(l) e^{-d l}$, {\it
etc.\/}, where $l$ is the smallest of the three cutoffs $l_x$,
$l_\perp$, $l_\omega$ to be found from equations $K(l_x) \Lambda^4 =
q_x^2 e^{2 l_x} Y(l_x)$, $\Lambda = q_\perp e^{l_\perp}$, and
$\omega_c^2 \omega = \omega_p^2(\Lambda e^{-l_\omega}) \nu_3(l_\omega)
\Lambda^2 e^{-(d + 2) l_\omega}$. These equations give rise to the
characteristic crossover  length and frequency scales $\xi_x = (K_0^5 /
Y_0)^{1/2} / (k_B T)^2$, $\xi_y = (K_0^3 / Y_0)^{1/2} / k_B T$, and
$\omega_y = (K_0 / \rho_0)^{1/2} \omega_p(\xi_y^{-1}) / \omega_c
\xi_y^2$.

Taking Eq.~(\ref{Y_K_l}) for the face value, we obtain $\gamma_Y =
\gamma_K = \frac12$ in 2D ($\epsilon = 1$) and discover the following
three types of asymptotic behavior:
\begin{eqnarray}
&& Y^R \sim Y_0 (\xi_y q_y)^{1/2},\quad
   K^R \sim K_0 (\xi_y q_y)^{-1/2},
\nonumber\\
&& \nu_3^R
   \sim \frac{K_0}{\rho_0 \omega_y \xi_y^2 (q_y \xi_y)^{3/4}},
\quad
   \nu_i^R \sim \frac{a_i Y_0}{\rho_0 \omega_y (q_y \xi_y)^{7/4}}
\label{limit_A}
\end{eqnarray}
for $q_x \ll \xi_x^{-1} (q_y \xi_y)^{3/2}$, $q_y \ll \xi_y^{-1}$,
$\omega \ll \omega_y (q_y \xi_y)^{9 / 4}$,
\begin{eqnarray}
&& Y^R \sim Y_0 (\xi_x q_x)^{1/3},\quad
   K^R \sim K_0 (\xi_x q_x)^{-1/3},
\nonumber\\
&& \nu_3^R \sim
           \frac{K_0}{\rho_0 \omega_y \xi_y^2 (q_x \xi_x)^{1/2}},
\quad
   \nu_i^R \sim \frac{a_i Y_0}{\rho_0 \omega_y (q_x \xi_x)^{7/6}}
\label{limit_B}
\end{eqnarray}
for $q_x \ll \xi_x^{-1}$, $q_y \ll \xi_y^{-1} (q_x \xi_x)^{2/3}$,
$\omega \ll \omega_y (q_x \xi_x)^{3 / 2}$, and
\begin{equation}
\nu_3^R \sim \frac{K_0}{\rho_0 \xi_y^2} \omega_y^{-2/3}
                \omega^{-1/3},
\quad
   \nu_i^R \sim \frac{a_i Y_0}{\rho_0 \omega_y^{2/9} \omega^{7/9}}
\label{limit_C}
\end{equation}
for $\omega \gg \omega_y (q_y \xi_y)^{9 / 4}$, $\omega \gg \omega_y (q_x
\xi_x)^{3 / 2}$. Here $a_1 = (1 - r)^2$, $a_4 = r^2$, $a_5 = (1 - r) r$,
and $r = \rho_0 C_0 / Y_0$. Thus, $a_1 a_4 = a_5^2$, which entails
the relation~\cite{Mazenko_83} $\Delta\eta_1 \Delta\eta_4 =
(\Delta\eta_5)^2$ derived earlier for the conventional smectics.

{\it Discussion.\/} Our one-loop-level results for $Y^R$ and $K^R$ turn
out to be exact~\cite{Golubovic_92} for the static limit
[Eqs.~(\ref{limit_A}) and (\ref{limit_B})]. It is possible
that this success transcends to the dynamics, in which case
the scaling exponents for viscosities in
Eqs.~(\ref{limit_A}--\ref{limit_C}) would be exact as well.

Within the domain of anomalous hydrodynamics summarized by
Eqs.~(\ref{limit_A}--\ref{limit_C}), the magnetophonon dispersion
relation becomes
\begin{eqnarray}
&& \displaystyle {\rm Re}\,\omega_m({\bf q}) \sim s c^{7/6}
(\xi_x q)^{5/3} \frac{\omega_p(\xi_x^{-1})}{\omega_c \xi_x}
\sqrt{\frac{Y_0}{\rho_0}},
\label{omega_m_R}\\
&& \displaystyle {\rm Im}\,\omega_m({\bf q}) \sim
s^2 c^{5/6} (\xi_x q)^{11/6}
(1 - 2 r)^2 \frac{\omega_y}{\xi_y} \sqrt{\frac{K_0}{Y_0}}
\label{alpha_m_R}
\end{eqnarray}
for a finite fixed $\theta$ and $q \ll \xi_x^{-1}$. In contrast to the
case of conventional smectics~\cite{Mazenko_83,Kats_Lebedev}, the
scaling of ${\rm Im}\,\omega_m({\bf q})$ is cut off by $l_x$
[Eq.~(\ref{limit_B})] not $l_\omega$ [Eq.~(\ref{limit_C})].

Strictly speaking, the smectic behavior in 2D can persist only up to
the length scale smaller than the average distance $\xi_d \sim
\exp(E_d / 2 k_B T)$ between dislocations, $E_d \sim K_0$ being the
dislocation energy~\cite{Toner_81}. The anomalous hydrodynamics can be
observed if $\xi_x < \xi_d$, which is satisfied at low $T$
where $\xi_d$ is exponentially large~\cite{Comment_on_nematic}.

In conclusion, we formulated a novel long-wavelength low-frequency
effective theory of the stripe phase arising in a two-dimensional
electron liquid near half-integral fillings of high Landau levels. Our
theory applies at relatively high temperatures and in the clean limit.
We demonstrated that the collective mode properties of the system
exhibit nontrivial power-law scaling, which can be verified by
microwave or surface acoustic wave measurements at a finite wave
vector or by inelastic light scattering experiments. Our future plans
include applying the proposed hydrodynamic approach to the problem of
the magnetotransport.

{\it Acknowledgements\/}.--- One of us (M.~F.) is grateful to
B.~I.~Shklovskii for stimulating his interest to hydrodynamic problems
and also to S.~Simon and L.~Pryadko for comments on the manuscript.
This work is supported by the U.S. Department of Energy under grants
Nos.~DE-FG02-90ER40542 and W-31-109-ENG-38, and also by
the NSF under grant No. DMR-91-20000.

\end{multicols}

\begin{references}
\bibitem{Comment_on_exchange} The parameters $Y$, $K$, and $\chi$
of the effective Hamiltonian are determined by the short-range
physics, which involves the competition between Hartree, exchange,
correlation, and the entropy parts of the free energy and can be
estimated following Ref.~\onlinecite{Koulakov_96}.
\bibitem{Comment_on_nematic} Unless
dislocations are generated quantum-mechanically even at
$T = 0$, leading to a ``quantum nematic'' phase~\cite{Fradkin_99}.
\end{references}
\end{document}